%% file: paper.tex
\documentclass{webofc}
\usepackage[varg]{txfonts}   

\usepackage{booktabs}
\usepackage{comment}

\newcommand{\rel}{\ensuremath{\mathrm{rel}}}

\newcommand{\wom}{\ensuremath{W_1^\mathrm{M}}\xspace}
\newcommand{\wop}{\ensuremath{W_1^\mathrm{P}}\xspace}
\newcommand{\woefp}{\ensuremath{W_1^\mathrm{EFP}}\xspace}
\newcommand{\etarel}{\ensuremath{\eta^{\rel}}\xspace}
\newcommand{\phirel}{\ensuremath{\phi^{\rel}}\xspace}
\newcommand{\mrel}{\ensuremath{m^{\rel}}\xspace}

\newcommand{\ptjet}{\ensuremath{{p}_{\mathrm{T,jet}}}\xspace}
\newcommand{\ptrel}{\ensuremath{{p}_{\mathrm{T}}^{\rel}}\xspace}

\newcommand{\JN}{\texttt{JetNet}\xspace}

\newcommand{\TeV}{\ensuremath{\,\text{Te\hspace{-.08em}V}}\xspace}

\usepackage{orcidlink}
 \usepackage{subcaption}
\usepackage{floatrow}
\usepackage{xcolor}
\definecolor{msg}{rgb}{0.59,0.45,0.65}
\definecolor{ngbr}{rgb}{0.65,0.55,0.0}
\definecolor{x4}{rgb}{0.51,0.70,0.40}

\usepackage{listings}
\usepackage{placeins}
\begin{document}
\title{DeepTreeGAN: Fast Generation of High Dimensional Point Clouds}

\author{
     \firstname{Moritz A.W. }\lastname{Scham}
     \inst{1,2,3}\fnsep\thanks{\email{moritz.scham@desy.de}} \and
     \firstname{Dirk} \lastname{Krücker}
     \inst{1}\and
     \firstname{Benno} \lastname{Käch}
     \inst{1,4}\and 
     \firstname{Kerstin} \lastname{Borras}
     \inst{1,3}
}

\institute{
     Deutsches Elektronen-Synchrotron DESY, Germany\and
     J\"ulich Supercomputing Centre, Institute for Advanced Simulation, Germany \and
     RWTH Aachen University, III. Physikalisches Institut A, Germany \and
     Universität Hamburg, Institut für Experimentalphysik, Germany
     }

\abstract{
In High Energy Physics, detailed and time-consuming simulations are used for particle interactions with detectors. To bypass these simulations with a generative model, the generation of large point clouds in a short time is required, while the complex dependencies between the particles must be correctly modelled. Particle showers are inherently tree-based processes, as each particle is produced by the decay or detector interaction of a particle of the previous generation.
In this work, we present a novel Graph Neural Network model (DeepTreeGAN) that is able to generate such point clouds in a tree-based manner. We show that this model can reproduce complex distributions, and we evaluate its performance on the public JetNet dataset.
}

\maketitle

\input{parts/intro}
\input{parts/architecture}
\input{parts/results}
\section{Conclusion}
This paper presents a new model (DeepTreeGAN) for generating large point clouds using a novel upscaling method for point clouds. It successfully replicates the complex distributions of the \JN datasets with high fidelity (Section~\ref{sec:resul}) and demonstrating its suitability for generating high-dimensional point clouds. The model's tree-based structure promises good scaling behavior for even larger point clouds, such as particle showers in high-granularity calorimeters. However, the preliminary version presented here uses the MPGAN critic that scales with $\mathcal{O}(\mathrm{points}^2)$ and thus, requires a new critic with better scaling behavior. In addition, the generator will need to be able to generate a variable number of points for each event, especially for the \JN datasets with up to 150 particles and the calorimeter use case. This will be the subject of future publications.


\section*{Acknowledgement}
\begin{acknowledgement}
Moritz Scham is funded by Helmholtz Association's Initiative and Networking Fund through Helmholtz AI (grant number: ZT-I-PF-5-3). Benno Käch is funded by Helmholtz Association's Initiative and Networking Fund through Helmholtz AI (grant number: ZT-I-PF-5-64).
This research was supported in part through the Maxwell computational resources operated at Deutsches Elektronen-Synchrotron DESY (Hamburg, Germany). The
authors acknowledge support from Deutsches Elektronen-Synchrotron DESY (Hamburg, Germany), a member of the Helmholtz Association HGF.
\end{acknowledgement}

\bibliography{paper}

\end{document}

%% file: parts/intro.tex
\section{Introduction}\label{intro}

In particle physics, detailed, near-perfect simulations of the underlying physical processes, the measurements and the details of the experimental apparatus are state of the art. However, accurate simulations of large and complex detectors, especially calorimeters, using current Monte Carlo-based tools such as those implemented in the Geant4~\cite{Agostinelli2003} toolkit
are computationally intensive. Currently, more than half of the worldwide Large Hadron Collider
(LHC) grid resources are used for the generation and processing of simulated data~\cite{Albrecht2019}. 
For the future High-Luminosity upgrade of the LHC (HL-LHC~\cite{Apollinari2015}), the computational requirements will exceed the computational resources without significant speedups, e.g.\ for the CMS experiment by 2028~\cite{Software2022} projecting the current technologies.
This will become even more demanding for future high-granularity calorimeters, e.g. the CMS HGCAL~\cite{CMS2017} with its complex geometry and extremely large number of channels.
To address these challenges, fast simulation approaches based on Deep Learning, including Generative Adversarial Networks (GAN)~\cite{Goodfellow2020} and Variational Autoencoders (VAE)~\cite{Kingma2013}, have been explored early~\cite{Oliveira2017,Paganini2018,Paganini2018a,Erdmann2018,Erdmann2019,Carminati2018} and deployed recently~\cite{Aad2022}.

The majority of these approaches represent the data as voxels living on three-dimensional grids. But especially for high-granularity calorimeters like the HGCAL, the calorimeter cells are highly irregular and the data is often sparse. In such a situation, it is beneficial to represent the data as Point Clouds (PC), e.g.\ tuples of cell energies and three-dimensional coordinates, i.e.\ four-dimensional point clouds~\cite{kaech2022jetflow,kaech2022point,schnake2022}. For modelling calorimeter showers, which are cascades of particles, it seems natural to model the creation of such point clouds as a tree-like process. Such an approach can be implemented by Graph Neural Networks (GNN)~\cite{Scarselli2009} and Graph Convolutions~\cite{kipf2016}. For modelling PC data representing three-dimensional objects, a similar approach~\cite{Shu2019} called tree-GAN~\cite{Shu2019} exists. For our use case, especially for modelling the large dynamic range of the energy dimension, this approach is not sufficient.
Here we report on the development of a largely extended Graph GAN approach, which we named {\it DeepTreeGAN} that can be applied to multiple areas of PC generation. To demonstrate the abilities of our model, we present a first application to the \JN dataset introduced in the next section. 

\subsection{JetNet and MPGAN}\label{sec:jetnet}

In this study, the \JN~\cite{Kansal2022} datasets are used. 
Each dataset contains PYTHIA~\cite{Sjostrand2008} jets with an energy of about 1\TeV, with each jet containing up to 30 or 150 constituents (here: 30). The datasets differ in the jet-initiating partons. Here, datasets for top quark, light quarks and gluon-initiated jets are studied~\cite{JetNet30v2,JetNet150v2}. Each dataset contains about 170k individual jets split as 110k/10k/50k for training/test/validation where the validation dataset is used for our results. The jet constituents, the particles, are clustered with a cone radius of $R = 0.8$. These particles are considered to be massless and can therefore be described by their 3-momenta or equivalently by transverse momentum $p_T$, pseudorapidity $\eta$, and azimuthal angle $\phi$. In the \JN datasets, these variables are given relative to the jet momentum: 
$\etarel_i \coloneqq \eta_i - \eta^\mathrm{jet}$,
$\phirel_i \coloneqq \phi_i - (\phi^\mathrm{jet} \mod 2\pi)$, and 
$p_{T,i}^\mathrm{rel} \coloneqq p_{T,i}/\ptjet$, where $i$ runs over the particles in a jet. 
Calculating the invariant mass from these relative quantities, for example, for the jet mass, implies
$\mrel = m_\mathrm{jet}/\ptjet$. 
The \JN library~\cite{Kansala} provides several metrics that are used in this study. Additionally, the authors provide a baseline model called MPGAN~\cite{Kansal}. 

%% file: parts/architecture.tex
\section{Architecture}\label{sec:arch}
\begin{figure}[th]
    \centering
    \includegraphics[width=0.9\textwidth,clip]{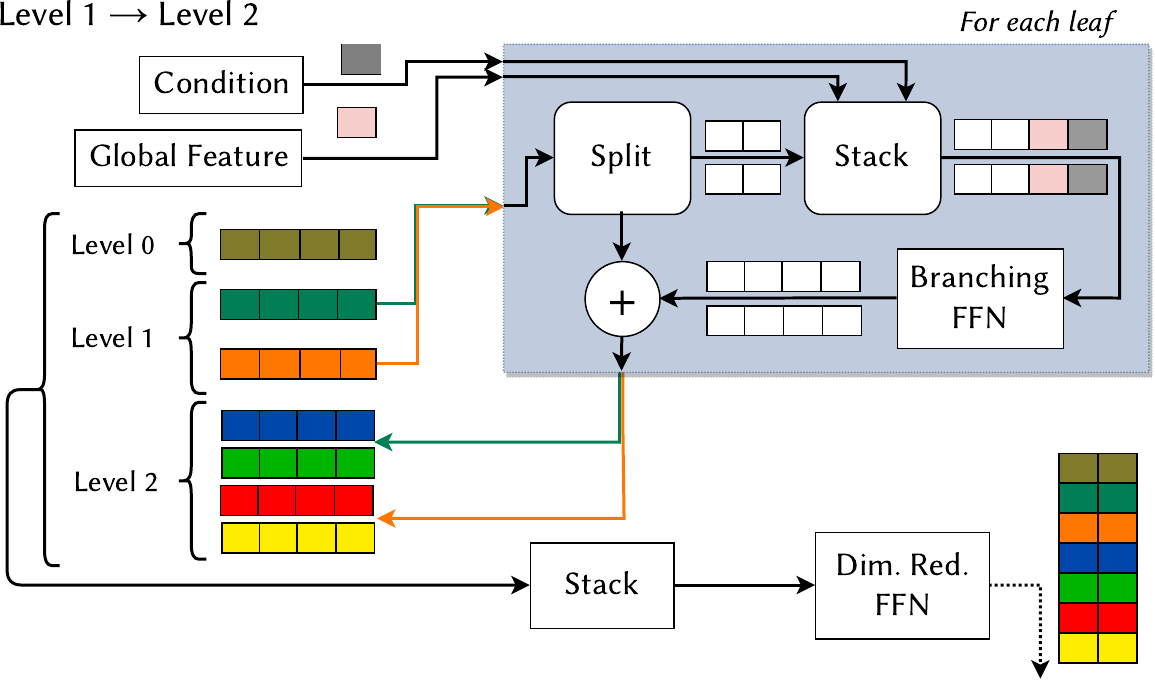}
    \caption{ \textbf{Branching Layer}. In this example, the nodes of the 2nd level of the tree are produced and attached as the new leaves.
    With the nodes from level 1 (dark green / orange) as the parents, the children in level 2 (blue+green / red+yellow) are generated independently:
    The vector of the parent is split into one part for each of the branches. To each of these vectors, the Global Feature (\ref{sec:arch}) and the global condition vector (here: number of constituents) are concatenated. The \emph{Branching FFN} maps the vectors to the same size as their parents. Then the feature vector of the parents is added to each of its children. 
    With the new children added as leaves to the tree, all the levels of the tree are stacked up and passed through a \emph{Dimensionality Reduction FFN}. With each Branching Layer, the number of nodes increases (1 $\to$ 2 $\to$  2*3 $\to$  2*3*5) and the number of features decreases (64 $\to$  33 $\to$  20 $\to$ 3).} 
    \label{fig:branching}
\end{figure}

The generator of this GAN constructs point clouds by starting with a random vector as the root of a tree and then attaching one level of leaves after the other. The output of the generator is the last level of the tree. 
Starting with the root node, a \emph{Branching Layer} (Figure~\ref{fig:branching}) takes the current leaves from the tree, maps each leaf to a given number nodes $n_i$ and attaches these nodes as new leaves to the tree. Then further \emph{Branching Layers} are applied until the desired number $\prod_i \mathrm{n}_i$ of nodes (here: 30 nodes) is reached. For these projections, multiple feed-forward neural networks (FFNs) are used.


\begin{figure}[ht]
        \floatbox[{\capbeside\thisfloatsetup{capbesideposition={right,top},capbesidewidth=0.6\textwidth}}]{figure}[\FBwidth]
    {\caption{\textbf{Ancestor Message Passing Layer}. The edges in the graph are constructed so that each node receives messages from each of its ancestors as well as itself. In the given example node 4 is updated with the aggregate of the messages from nodes 1, 2 and 4. As a message-passing algorithm, GINConv~\cite{xu2019powerful}  -- as implemented in PyTorch Geometric \cite{pyg} -- is chosen. For GINConv, the messages are the features of the source node (here: the ancestor). These messages are aggregated by summing over all messages addressed to the target node. These aggregates are added to the feature vector of the target node (scaled with a learnable weight) and passed through an FFN. 
    The nodes are then updated with the output of the FFNs plus, as a residual connection, the original feature vector of the nodes themselves.
    }\label{fig:acmpl}}%
    {\includegraphics[keepaspectratio,width=0.4\textwidth]{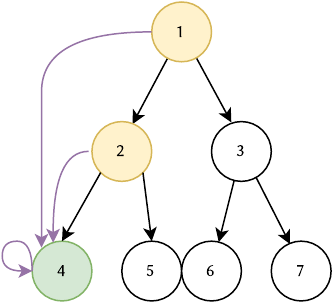}}
    
\end{figure}

In between the \emph{Branching Layers} a Message Passing Layer (MPL) passes the information down the tree. In this step, each of the nodes is updated with information from its ancestors. This \emph{Ancestor MPL} is described in Figure~\ref{fig:acmpl}.

A \emph{Global Feature Layer} is introduced that condenses the information of the current leaves in a single vector: The feature vectors of the leaves are passed through a first FFN, summed up and passed through a second FFN. The resulting vector is then used -- together with a global condition vector -- to condition the following Branching Layer and Ancestor MPL. The global condition vector contains only the number of constituents in the event.
For each of the levels of the tree, first the Global Feature Layer is executed, then the Branching Layer and then the Ancestor MPL. 

The FFNs of this model consist of 3 hidden layers of 100 nodes without bias. The first two hidden layers are followed by a Batch Normalization~\cite{batchnorm} layer and LeakyReLU activation with a negative slope of 0.1.
As a critic, the MPGAN critic~\cite{Kansal} is used.

\subsection{Training}
Generator and Critic are trained using the Hinge~\cite{Hinge} loss, with a ratio of 1 to 2 gradient steps.
The optimizer for the Generator is Adam with $\beta_1 = 0.9, \beta_2 = 0.999$, a weight decay of $10^{-4}$ and a learning rate of $10^{-5}$. The optimizer for the critic is Adam with $\beta_1 = 0.0,\beta_2 = 0.9$, a weight decay of $10^{-4}$ and a learning rate of $3*10^{-5}$.

\subsection{Comparison to  tree-GAN}

Compared to the model proposed in~\cite{Shu2019}, DeepTreeGAN introduces some major changes: The branching (Figure~\ref{fig:branching}) uses two FFNs instead of matrix multiplication: one for increasing the number of nodes, the other for reducing the number of features. As the new leaves have the same size as their ancestors, a single Dimensionality Reduction FFN can be applied to all nodes of the tree.
After the branching, both models implement a graph convolution step, passing information from the ancestors to their children. In tree-GAN, the leaves are updated using the leaf itself and its ancestors, each multiplied with a separate trainable matrix.
In DeepTreeGAN, all nodes in the tree have the same dimension at all times. Thus, all nodes can be updated at the same time, using a single FFN in the Ancestor MPL (Figure~\ref{fig:acmpl}).

\subsection{Preprocessing \& Postprocessing}
For the training, $\etarel$ and $\phirel$ are scaled separately to a normal distribution. 
The $\ptrel$ distribution is transformed using the Box-Cox transformation with standardizing implemented in \cite{Sklearn}. To produce samples, the inverse scaling is applied to the output of the generator.

%% file: parts/results.tex
\section{Results}\label{sec:resul}
\begin{figure}[ht]
    \centering
    \begin{subfigure}[b]{0.32\textwidth}
        \centering
        \includegraphics[keepaspectratio,width=.98\textwidth,trim={0 0.5cm 0 0},clip]{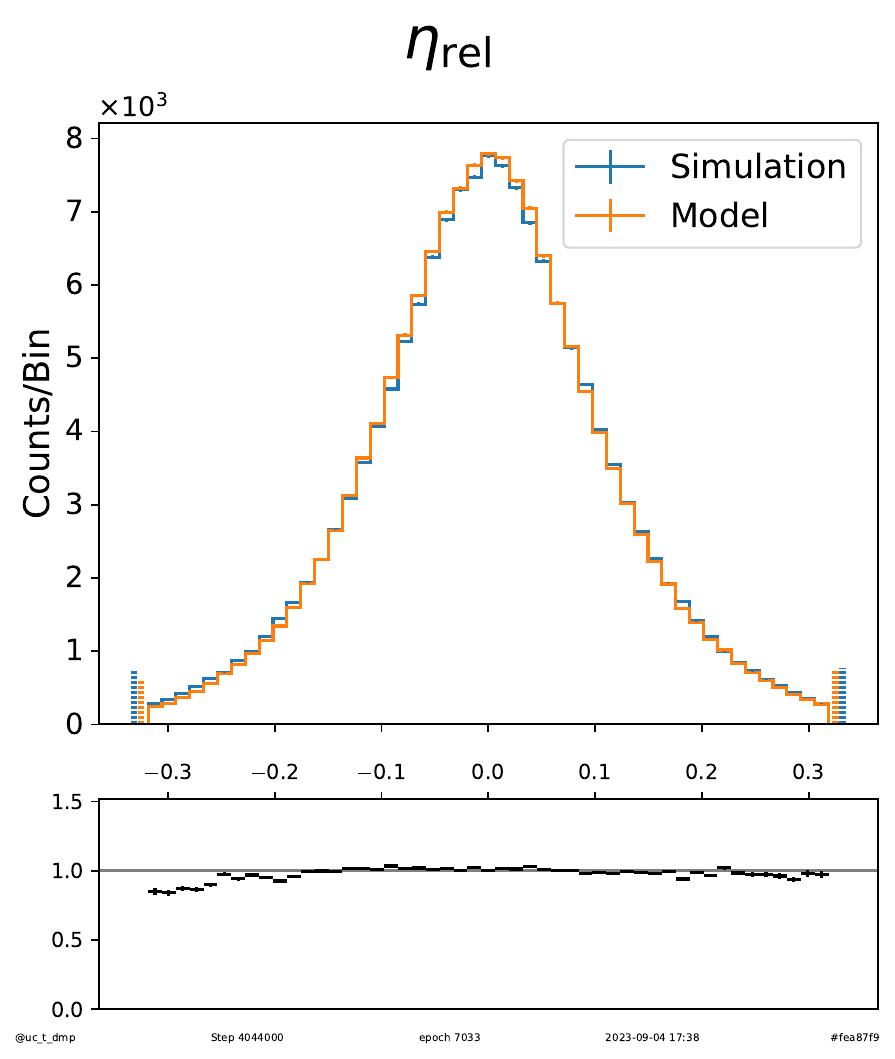}
    \end{subfigure}
        \begin{subfigure}[b]{0.32\textwidth}
        \centering
        \includegraphics[keepaspectratio,width=.98\textwidth,height=.6\textheight,trim={0 0.5cm 0 0},clip]{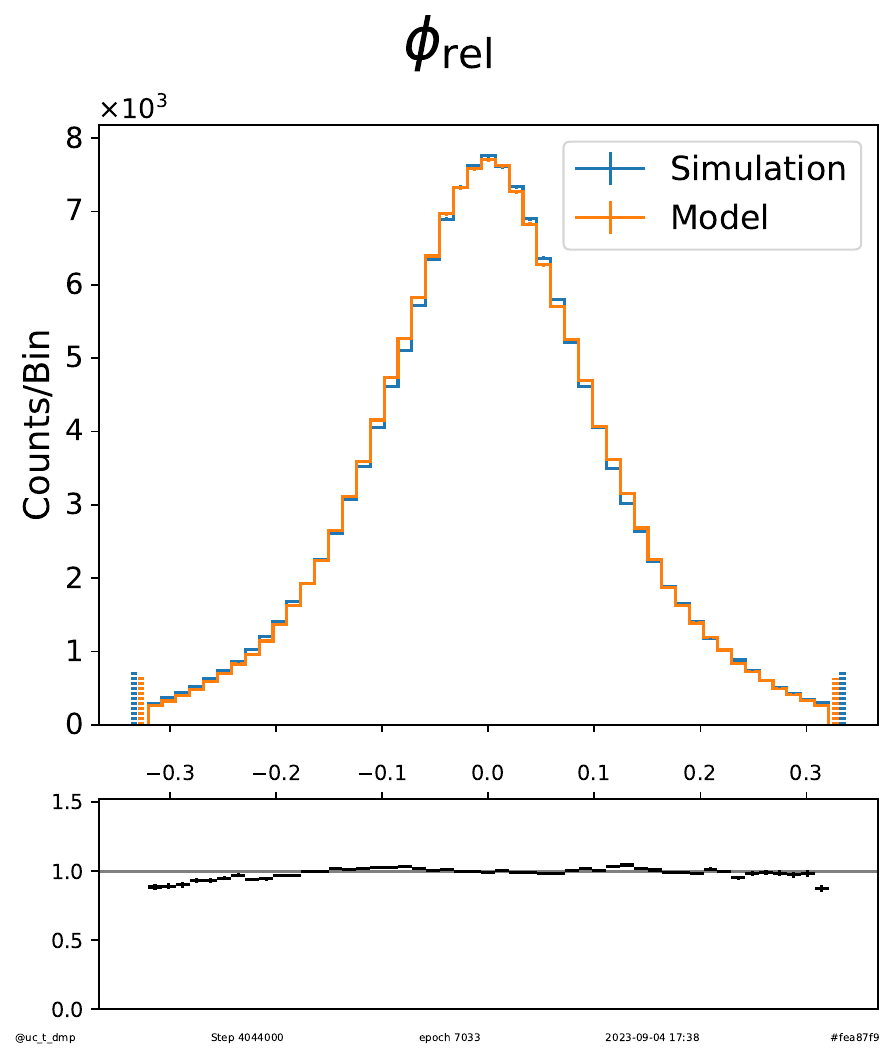}
    \end{subfigure}
        \begin{subfigure}[b]{0.32\textwidth}
        \centering
        \includegraphics[keepaspectratio,width=.98\textwidth,height=.6\textheight,trim={0 0.5cm 0 0},clip]{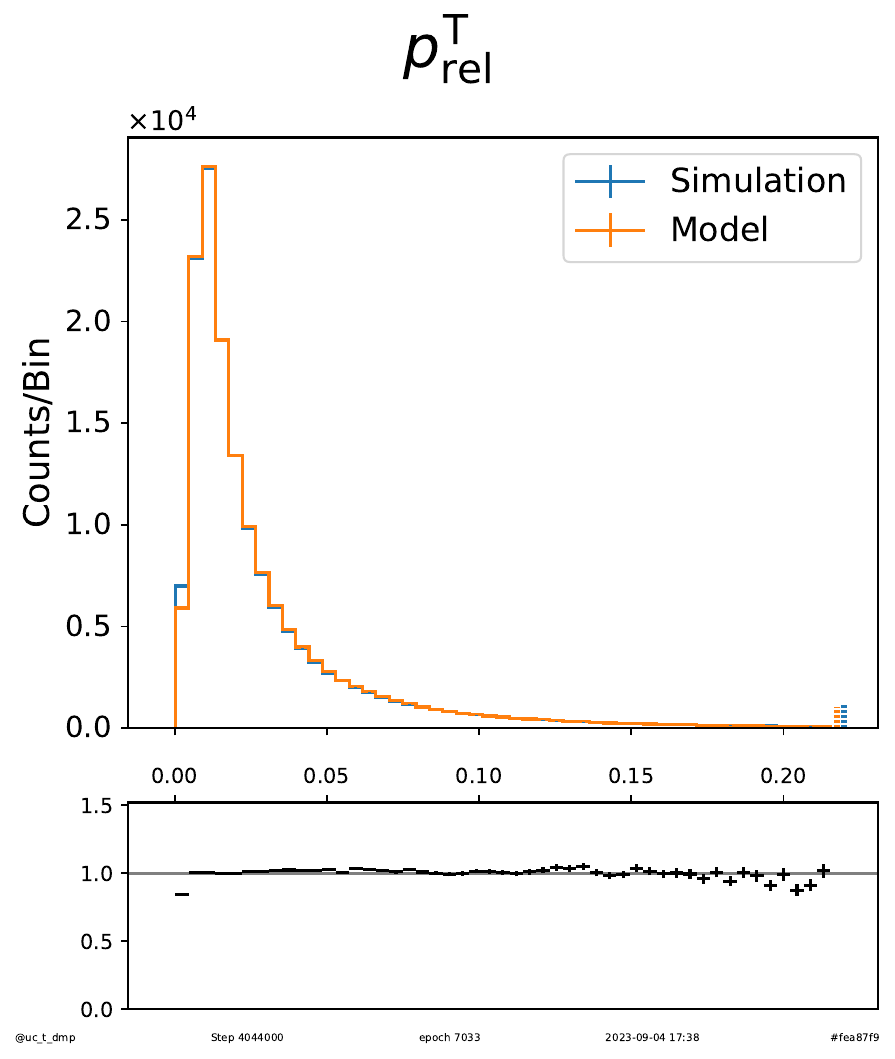}
    \end{subfigure}
    \caption{The global particle distributions for the top-quark test dataset, as described in section~\ref{sec:jetnet}.}\label{fig:t-mar}
\end{figure}
\begin{figure}
    \begin{subfigure}[b]{0.32\textwidth}
        \centering
        \includegraphics[keepaspectratio,width=.98\textwidth,trim={0 0.5cm 0 0},clip]{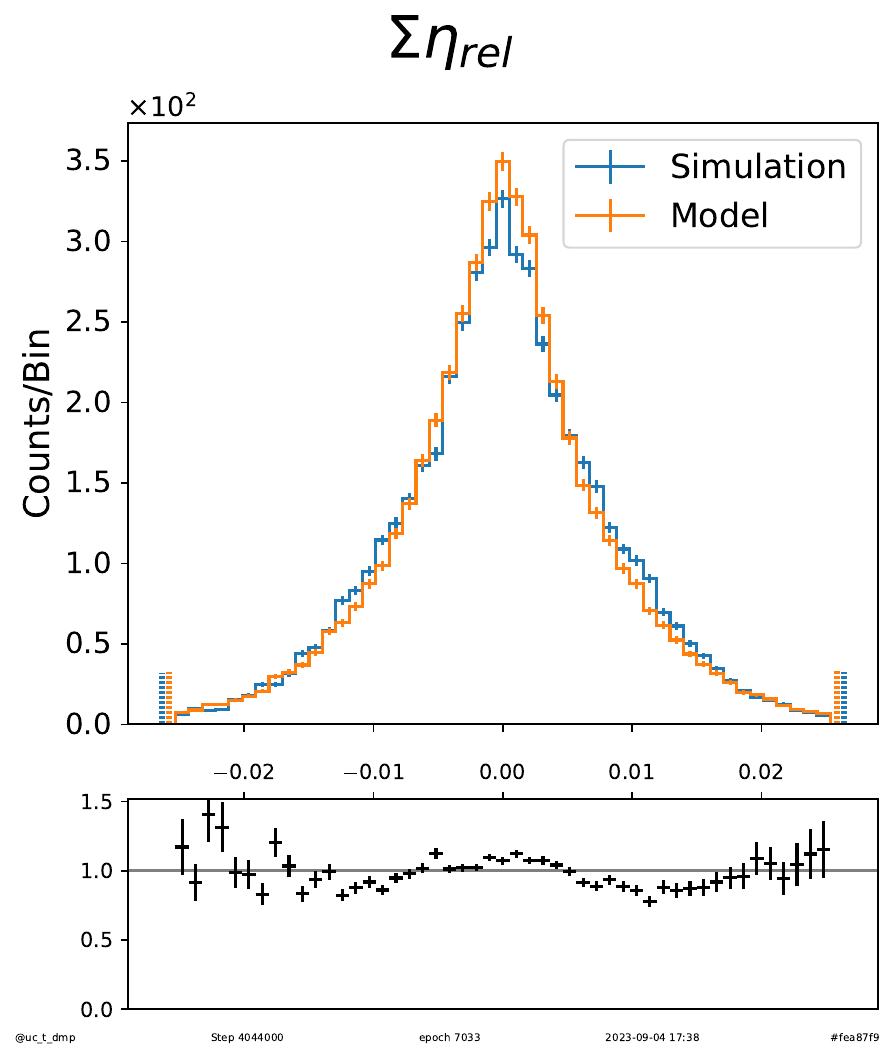}
    \end{subfigure}
        \begin{subfigure}[b]{0.32\textwidth}
        \centering
        \includegraphics[keepaspectratio,width=.98\textwidth,height=.6\textheight,trim={0 0.5cm 0 0},clip]{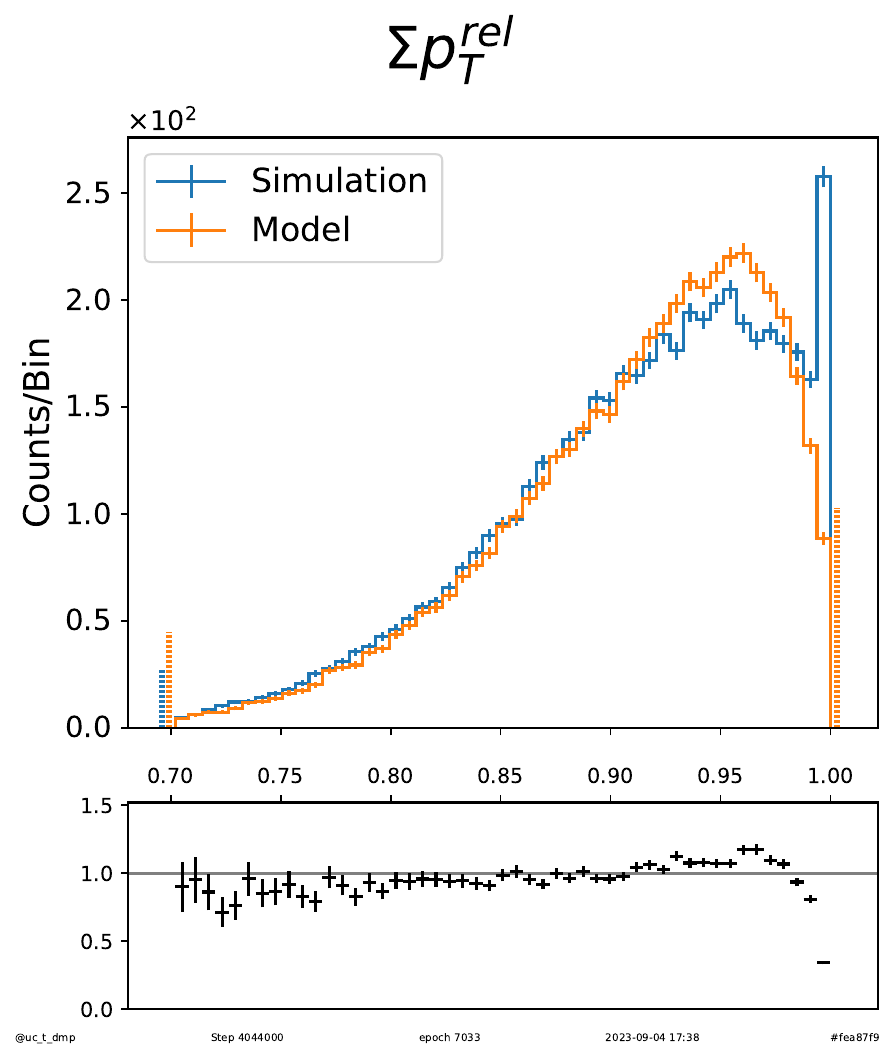}
    \end{subfigure}
        \begin{subfigure}[b]{0.32\textwidth}
        \centering
        \includegraphics[keepaspectratio,width=.98\textwidth,height=.6\textheight,trim={0 0.5cm 0 0},clip]{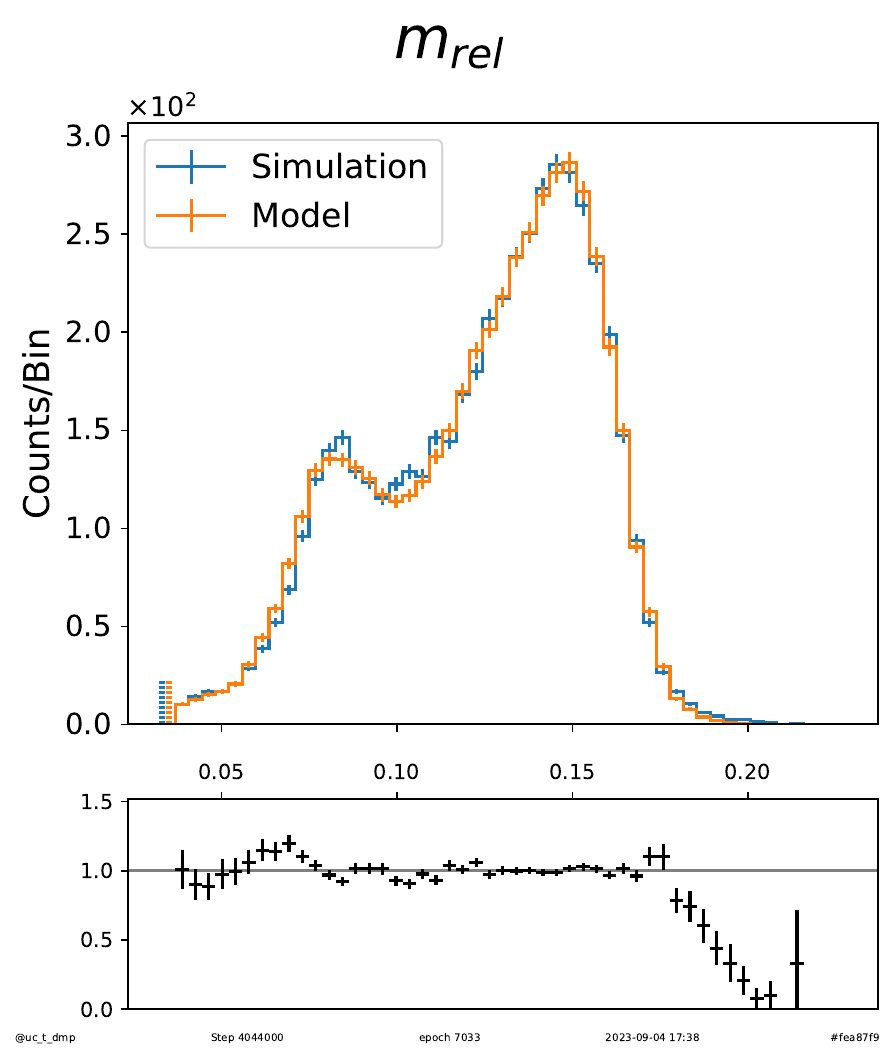}
    \end{subfigure}
    \caption{The distributions of the jet variables in the top-quark test dataset.}
    \label{fig:t-jet}
\end{figure}
The variables of the constituents of the top quark dataset are shown in Figure~\ref{fig:t-mar}. 
The generated distributions of the constituents closely match the distribution of the data. The jet variables shown in Figure~\ref{fig:t-jet} are significantly harder to model. Our model successfully reproduces the double peak structure in the relative jet mass, showing a significant advantage over tree-GAN as shown in \cite[Figure~3]{Kansal2022}.

\input{parts/metricstable}

In Table~\ref{tab:metrics} the achieved metrics are shown. While our model shows an advantage over the MPGAN baseline on the top quark dataset and an at least similar performance on the light quarks' dataset, the performance is worse for the gluon dataset. At the current stage, the generator always produces the full 30 constituents for each of the events. Thus, the performance for events with less than 30 constituents is limited, as is frequently the case for the gluon jets.
The models and code are available on GitHub\footnote{\texttt{\url{https://github.com/DeGeSim/chep23DeepTreeGAN}}}.

%% file: parts/metricstable.tex
\begin{table*}[bt]
    \caption{Comparison of the proposed DeepTreeGAN to the MPGAN baseline with metrics introduced in~\cite{Kansal2022}. Significant advantages for a model in a given metric are highlighted bold. The `Limit' row gives the measured in-sample distance by sampling datasets from the training dataset and calculating the metrics. Note that the MPGAN model was selected for \wom, and the testing sample was used to compute the given metrics. The $W_1$ metrics are calculated with 10000 samples, FPND is calculated with 50000 samples.} \label{tab:metrics}
    \begin{tabular}{llllll}
        \toprule
        Dataset      & Model    & $\wom \times10^{3}$ & $\wop \times10^{3}$ & $\woefp \times10^{5}$ & FPND            \\
        \midrule
        Gluon        & Limit    & $0.7\pm 0.2$           & $0.44\pm 0.09$         & $0.63\pm 0.07$           & $<10^{-3}$              \\
        ~            & MPGAN    & $\mathbf{0.7}\pm 0.2$  & $\mathbf{0.9}\pm 0.3$  & $\mathbf{0.7}\pm 0.2$    & $\mathbf{0.12}$ \\
        ~            & DeepTreeGAN & $1.8\pm 0.2$           & $1.6\pm 0.6$           & $2.1\pm 0.2$             & $0.34$          \\
        \midrule
        Light Quarks & Limit    & $0.5\pm 0.1$           & $0.5\pm 0.1$           & $0.46\pm 0.04$           & $<10^{-3}$               \\
        ~            & MPGAN    & $\mathbf{0.7}\pm 0.2$  & $4.9\pm 0.5$           & $0.7\pm 0.4$    & $0.35$          \\
        ~            & DeepTreeGAN & $0.9\pm 0.2$           & $\mathbf{1.7}\pm 0.3$  & $0.9\pm 0.2$             & $\mathbf{0.15}$ \\
        \midrule
        Top Quarks   & Limit    & $0.51\pm 0.07$         & $0.55\pm 0.07$         & $1.1\pm 0.1$             & $<10^{-3}$               \\
        ~            & MPGAN    & $0.6\pm 0.2$  & $2.3\pm 0.3$           & $2\pm 1$                 & $0.37$          \\
        ~            & DeepTreeGAN & $0.6\pm 0.1$           & $\mathbf{1.1}\pm 0.5$  & $\mathbf{1.3}\pm 0.3$                 & $\mathbf{0.14}$ \\
        \bottomrule
    \end{tabular}
\end{table*}